\documentclass[conference]{IEEEtran}
\IEEEoverridecommandlockouts

\usepackage{verbatim}
\usepackage{amssymb}
\usepackage{graphicx}
\usepackage{url}
\usepackage{epstopdf}
\usepackage{float}
\usepackage{makeidx}
\usepackage{graphicx}
\usepackage{wrapfig}
\usepackage{comment}
\usepackage{csquotes}
\usepackage{url}
\usepackage{xcolor,colortbl}
\usepackage{color}
\usepackage{fancyvrb}
\usepackage{epsfig}\restylefloat{table}
\usepackage{setspace}
\usepackage{multirow}
\usepackage[bookmarks,bookmarksopen,bookmarksdepth=2,hidelinks]{hyperref}
\usepackage{makeidx}
\usepackage{threeparttable}
\usepackage{mathtools}
\usepackage[ruled, linesnumbered, vlined]{algorithm2e}
\usepackage{algpseudocode}\algrenewcommand\textproc{}
\usepackage{listings}
\usepackage{amssymb}
\usepackage{listings}
\usepackage[numbers, sectionbib]{natbib} 
\usepackage[misc]{ifsym}
\usepackage[nolist]{acronym}
\usepackage{tikz}
\usepackage{flushend}
\usepackage{framed}
\SetAlFnt{\small}
\SetAlCapFnt{\small}
\SetAlCapNameFnt{\small}
\SetAlCapHSkip{0pt}
\IncMargin{-\parindent}

\makeatletter
\renewcommand\subsection{\@startsection{subsection}{2}{\z@}
                        {-8\p@ \@plus -4\p@ \@minus -4\p@}
                        {6\p@ \@plus 4\p@ \@minus 4\p@}
                        {\normalfont\normalsize\bfseries\boldmath
                        \rightskip=\z@ \@plus 8em\pretolerance=10000 }}
\renewcommand\subsubsection{\@startsection{subsubsection}{3}{\z@}
                        {-4\p@ \@plus -4\p@ \@minus -4\p@}
                        {-1.5em \@plus -0.22em \@minus -0.1em}
                        {\normalfont\normalsize\bfseries\boldmath}}
\makeatother

\makeindex

\begin{acronym}
\acro{RMA}{\emph{Remote Memory Access}}
\acro{RMO}{\emph{Remote Memory Operation}}
\acro{AMO}{\emph{Atomic Memory Operation}}
\acro{PE}{\emph{Processing Element}}
\acrodefplural{PE}[PEs]{\emph{Processing Elements}}
\acro{PGAS}{\emph{Partitioned Global Address Space}}
\acro{OSM14}{OpenSHMEM specification version 1.4}
\acro{OSHM}[OSHM]{OpenSHMEM specification}
\end{acronym}

\newcommand{\openshmem}[1][]{{Open\-SHMEM\ifthenelse{\equal{#1}{}}{}{~#1}}\xspace}

\usetikzlibrary{calc}
\makeatletter
\newlength{\mylength}
\xdef\CircleFactor{1.1}
\newsavebox{\mybox}
\newcommand*\circled[2][draw=blue]{\savebox\mybox{\vbox{\vphantom{WL1/}#1}}\setlength\mylength{\dimexpr\CircleFactor\dimexpr\ht\mybox+\dp\mybox\relax\relax}\tikzset{mystyle/.style={circle,#1,minimum height={\mylength}}}
\tikz[baseline=(char.base)]
\node[mystyle] (char) {#2};}
\makeatletter

\begin{document}

\IEEEoverridecommandlockouts
\IEEEpubid{\begin{minipage}[t]{\textwidth}\ \\[10pt]
     \begin{framed}
     \centering\normalsize{This work is planned for possible publication to the
     IEEE. Copyright may be transferred without notice, after which this version
     may no longer accessible.}
     \end{framed}
\end{minipage}}

\title{Exploring GPU Stream-Aware Message Passing using Triggered Operations}
\author{
    \IEEEauthorblockN{Naveen Namashivayam}
        \IEEEauthorblockA{\textit{Hewlett Packard Enterprise, USA}\\
                          naveen.ravi@hpe.com}
    \IEEEauthorblockN{Nick Radcliffe}
        \IEEEauthorblockA{\textit{Hewlett Packard Enterprise, USA}\\
                          nick.radcliffe@hpe.com}
    \and
    \IEEEauthorblockN{Krishna Kandalla}
        \IEEEauthorblockA{\textit{Hewlett Packard Enterprise, USA}\\
                          krishnachaitanya.kandalla@hpe.com}
    \IEEEauthorblockN{Larry Kaplan}
        \IEEEauthorblockA{\textit{Hewlett Packard Enterprise, USA}\\
                          larry.kaplan@hpe.com}
    \and
    \IEEEauthorblockN{Trey White}
        \IEEEauthorblockA{\textit{Hewlett Packard Enterprise, USA}\\
        trey.white@hpe.com}
    \IEEEauthorblockN{Mark Pagel}
        \IEEEauthorblockA{\textit{Hewlett Packard Enterprise, USA}\\
                          mark.pagel@hpe.com}
}

\maketitle

\begin{abstract}
Modern heterogeneous supercomputing systems are comprised of compute blades that
offer CPUs and GPUs. On such systems, it is essential to move data efficiently
between these different compute engines across a high-speed network. While
current generation scientific applications and systems software stacks are
\textit{GPU-aware}, CPU threads are still required to orchestrate data moving
communication operations and inter-process synchronization operations.

A new \textit{GPU stream-aware} MPI communication strategy called
\textit{stream-triggered (ST) communication} is explored to allow
offloading both computation and communication control paths to the GPU. The
proposed ST communication strategy is implemented on HPE Slingshot Interconnects
over a new proprietary HPE Slingshot NIC (Slingshot 11) using the supported
\textit{triggered operations} feature. Performance of the proposed new
communication strategy is evaluated using a microbenchmark kernel called Faces,
based on the nearest-neighbor communication pattern in the CORAL-2 Nekbone
benchmark, over a heterogeneous node architecture consisting of AMD CPUs and
GPUs.
\end{abstract}


\section{Introduction}\label{sec:introduction}

\IEEEPARstart{C}{urrent}-generation scientific applications and systems-software
stacks are using \textit{GPU-aware}~\cite{GPU-aware-MVAPICH} Message Passing
Interface (MPI)~\cite{mpi} implementations. GPU-awareness for inter-node MPI
data movement using Remote Direct Memory Access
(RDMA)~\cite{GPU-aware-MVAPICH-inter-node,GPU-aware-MVAPICH-inter-node-2} allows
buffers to directly move from the GPU memory to a network adapter without
staging through host memory. For intra-node MPI transfers, buffer move
through various Peer-to-Peer (P2P) data transfer
mechanisms~\cite{GPU-aware-intra-node} supported by different GPU vendors.

Even with such GPU-awareness in the MPI software stack, CPU threads are still
required to orchestrate the data-moving communication and inter-process
synchronization operations. This requirement results in all communication and
synchronization operations occurring at GPU kernel boundaries.

\begin{figure}[ht]
  \includegraphics[width=\linewidth]{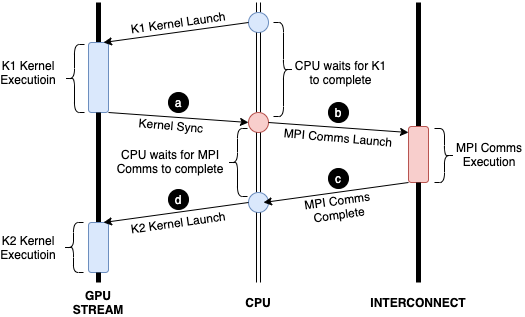}
  \caption{Illustrating sequence of events on a typical GPU-aware parallel
  application that relies on MPI for inter-process communication and
  synchronization operations.}
  \label{fig:normal-execution}
\end{figure}

Fig.~\ref{fig:normal-execution} demonstrates the sequence of events for a
typical GPU-aware parallel application that relies on MPI for inter-process
communication and synchronization operations. An MPI process running on the CPU 
\begin{tiny}\circled[text=white,fill=black,draw=black]{a}\end{tiny} first
synchronizes with the local GPU device to ensure completion of the compute
kernel (K1) execution. Next, it
\begin{tiny}\circled[text=white,fill=black,draw=black]{b}\end{tiny},
launches, progresses, and
\begin{tiny}\circled[text=white,fill=black,draw=black]{c}\end{tiny} completes
the inter-process communication/synchronization operations. Subsequent compute
kernels (K2) on the GPU are
\begin{tiny}\circled[text=white,fill=black,draw=black]{d}\end{tiny}
launched only after the inter-process communication operations have completed.
This behavior creates potentially expensive synchronization points at kernel
boundaries that require the CPU to synchronize with the GPU and Network
Interface Controller (NIC) devices.

A \textit{GPU stream}~\cite{cuda-stream} is a queue of device operations.
GPU compute kernel concurrency is achieved through creating multiple concurrent
streams. Operations issued on a stream get executed in the order in which these
operations were issued. Also, the execution of these operations is
asynchronous with respect to operations in other streams.
A new \textit{GPU stream-aware} MPI communication strategy called
\textit{stream-triggered (ST) communication} is proposed and explored to offload
both the computation and communication control paths to the GPU.

The new proposed communication scheme allows the CPU to create network command
descriptors with \textit{deferred execution} semantics and append them to the
NIC command queue. These command descriptors have special attributes that allow
them to get \textit{triggered}~\cite{tops-portal} at a later point in time when
certain conditions are reached. In addition, the CPU also creates control
operations and appends them to the GPU stream. These operations will be executed
by the GPU control processor in sequential order, relative to other operations
enqueued in the GPU stream. When these control operations are executed by the
GPU control processor, they act as \textit{triggers} that initiate the execution
of the previously appended network command descriptors in the NIC's command
queue.

This approach also allows the GPU control processor to synchronize with the NIC
to determine the successful completion of communication operations. This
synchronization step involves the use of hardware counters in the NIC. Thus, ST
tries to minimize the need for synchronization between the CPU and GPU.

\subsection{Contributions of This Work}
\label{sec:contrib}

The following are the major contributions.

\begin{enumerate}
    \item Propose a GPU stream-aware MPI communication strategy to offload both
    computation and communication control paths to the GPU;
    \item Implement the proposed communication strategy on HPE Slingshot
    Interconnects exploiting the \textit{triggered operations}
    feature~\cite{tops-portal} available in the new HPE Slingshot
    NIC (Slingshot 11)~\cite{slingshot1,slingshot2,slingshot};
    \item To demonstrate the effectiveness of the new proposed communication
    strategy, create a microbenchmark kernel called Faces based on the
    nearest-neighbor communication pattern in the CORAL-2~\cite{coral2-bm}
    Nekbone benchmark~\cite{nekbone-bm}; and
    \item Experiment with proposed solutions on a heterogeneous system
    architecture with support for AMD-based CPU and GPU processors.
\end{enumerate}

\section{Background}
\label{src:bground}

This section provides an overview of the ST communication
strategy in MPI, along with the \textit{deferred execution} features provided by
the HPE Slingshot NIC (Slingshot 11) to support the proposed ST interface.

\subsection{MPI Control and Data Paths}
\label{src:mpi-control-data-paths}

MPI communication operations in GPU-aware applications are typically comprised
of \textit{control paths} and \textit{data paths}. Control paths correspond to
coordination operations that occur between the application process running on
the CPU, the control processor on the GPU device, application compute kernels
running on the GPU device, and the NIC. Data paths refer to
those operations that involve moving data between CPU-attached and GPU-attached
memory regions. These data movement operations can occur within the same compute
node or between different compute nodes across a high-speed network.
The data paths are typically handled by the NIC (for inter-node data transfer
operations) and the GPU DMA engines (for intra-node peer-to-peer data transfer
operations).

As detailed in Fig.~\ref{fig:normal-execution}, applications typically
experience expensive synchronization points, and an application process running
on the CPU is closely involved in the progress of control paths.

Fig.~\ref{fig:st-execution} includes a high-level description of the proposed
new ST communication strategy. This strategy enables a GPU-aware application to
offload control paths to underlying implementation and hardware components.

\subsection{Stream Triggered Communication Overview}
\label{src:st-bground}

A parallel application using the ST strategy continues to manage compute kernels
on the GPU via existing mechanisms. In addition, the ST strategy allows an
application process running on the CPU to define a set of GPU stream based MPI
communication operations. The new proposed MPI operations are detailed in
Section~\ref{src:st-mpi-interface}.

These communication operations can be scheduled for execution at a later point
in time. More importantly, in addition to offering a deferred execution model,
the ST strategy enables the GPU control processor (GPU CP) to be closely
involved in the control paths associated with MPI communication operations.
Along with managing compute kernels defined by the application, the GPU CP
coordinates with the NIC to manage the control paths of MPI communication
operations. This design minimizes the need for the application process
running on the CPU to drive the control paths of MPI communication operations.

\begin{figure}[ht]
  \includegraphics[width=\linewidth]{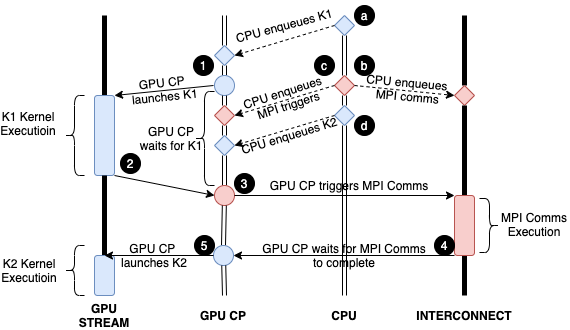}
  \caption{Illustrating the sequence of events on a parallel application that
  uses ST MPI execution for inter-process communication and
  synchronization.}
  \label{fig:st-execution}
\end{figure}

As shown in the Fig.~\ref{fig:st-execution}, an application process running on
the CPU enqueues
\begin{tiny}\circled[text=white,fill=black,draw=black]{a}\end{tiny} GPU
kernel K1 to the GPU stream,
\begin{tiny}\circled[text=white,fill=black,draw=black]{b}\end{tiny} triggered
ST-based MPI operations to the NIC,
\begin{tiny}\circled[text=white,fill=black,draw=black]{c}\end{tiny} the
corresponding trigger operations to the GPU stream, and
\begin{tiny}\circled[text=white,fill=black,draw=black]{d}\end{tiny} GPU kernel
K2 to the same stream. In Fig.~\ref{fig:st-execution}, the enqueue operations
from the CPU are represented via dashed arrows, and the actual executions of
various operations from the GPU Control Processor are represented as solid
arrows.

\begin{tiny}\circled[text=white,fill=black,draw=black]{1}\end{tiny} The GPU CP
launches K1 and
\begin{tiny}\circled[text=white,fill=black,draw=black]{2}\end{tiny} waits for it
to complete. Once K1 completes,
\begin{tiny}\circled[text=white,fill=black,draw=black]{3}\end{tiny} the GPU
CP triggers the execution of MPI operations and
\begin{tiny}\circled[text=white,fill=black,draw=black]{4}\end{tiny} waits for
these operations to finish. Next,
\begin{tiny}\circled[text=white,fill=black,draw=black]{5}\end{tiny} the GPU
CP launches K2.

Thus, an application process running on the CPU enqueues operations to the NIC
command queue and GPU stream, but the CPU is not directly involved in the
control paths of MPI communication operations and subsequent kernel launch and
tear-down operations. The CPU also does not directly wait for MPI communication
operations to complete. The GPU CP manages the control paths, and this
potentially eliminates the expensive synchronization points between an
application process and its GPU device.

\subsection{Slingshot 11 Triggered Operations using Libfabric Deferred Work
Queues}
\label{src:tops-ss11}

Triggered operations~\cite{tops-portal} supported by Slingshot 11 are the key
hardware features required for implementing the proposed ST MPI communication
scheme. They are exposed using the deferred work queue (DWQ)~\cite{fi_trigger}
features in the Libfabric~\cite{libfabrics} SW stack. DWQ allows an application
to enqueue a list of operation with deferred execution semantics. Each DWQ
descriptor is comprised of a DMA descriptor, a trigger counter object, a
completion counter object, and a trigger threshold value.

When the MPI implementation submits a DWQ send operation to the NIC command
queue, this operation is not executed by the NIC immediately. The operation gets
executed when the associated trigger counter object in the DWQ operation reaches
a required threshold value. The GPU CP is responsible for performing this
update. Similarly, the completion of the DWQ operation is monitored by the
GPU CP using the completion counter object which can be used for performing
synchronizations or execute other enqueued DWQ operations.

The current generation Slingshot 11 NIC provides support for the following DWQ
operations: (1) tagged and untagged sends, (2) one-sided RMA operations, and
(3) fetching and non-fetching atomic operations. Missing DWQ features required
by the MPI ST semantics, like the tagged and untagged receive operations, are
emulated with an internal asynchronous progress thread per MPI process.

The trigger and completion counter NIC objects can be mapped to GPU CP
accessible memory pointers for the GPU CP to directly access these NIC
counter objects.

\subsection{Stream Memory Operations}
\label{src:stream-memory-operations}

GPU stream memory operations are used to support the proposed ST communication
operations. Specifically, they are used to update and monitor the trigger and
completion counter objects specified in Section.~\ref{src:tops-ss11}. There are
two common stream memory operations supported by different GPU vendors:
(1) \textit{writeValue} and (2) \textit{waitValue}.

A \textit{writeValue} operation enqueues a write command to the GPU stream, and
the write operation is performed only after all earlier commands to this stream
have completed execution. A \textit{waitValue} operation enqueues a wait
command to the GPU stream. All operations enqueued on this stream after the
wait command will not be executed until the defined wait condition is satisfied.

The function prototypes of the \textit{hipStreamWriteValue64} and
\textit{hipStreamWaitValue64} operations on the AMD HIP
runtime~\cite{hip-stream-memory-ops} are shown in
Fig.~\ref{fig:hip-stream-memory-ops}. For brevity, similar stream memory
operations available in the NVIDIA CUDA runtime~\cite{cuda-stream-memory-ops}
are not shown.

\begin{figure}[!ht]
\begin{minipage}{\linewidth}
\lstset{
    language=C++,
    basicstyle=\tt\small,
    frame = single,
    morekeywords={int64_t, hipStream_t, hipError_t, uint64_t, unsigned int},
}
\begin{lstlisting}
hipError_t hipStreamWriteValue64(
    hipStream_t stream, void* ptr, int64_t val);

hipError_t hipStreamWaitValue64(
    hipStream_t stream, void* ptr, int64_t val,
    uint64_t mask, unsigned int flags);
\end{lstlisting}
\end{minipage}
\caption{Function prototypes for AMD HIP stream memory operations.}
\label{fig:hip-stream-memory-ops}
\end{figure}

\subsection{Mapping Stream Memory and SS11 DWQ Operations}
The \textit{writeValue} and \textit{waitValue} operations are used by mapping
the trigger and completion counter objects to GPU accessible memory
pointers. In brief, the GPU CP using the enqueued \textit{writeValue} operation
writes to the trigger counter object based on the threshold value. This
write operation acts as a trigger for the previously enqueued
DWQ(Section~\ref{src:tops-ss11}) operation to be executed. Similarly, the GPU CP using
the enqueued \textit{waitValue} operation monitors the completion counter object
associated with the DWQ operation to wait for its completion.

\section{Proposed ST MPI Interface}
\label{src:st-mpi-interface}

This section provides a brief description of the proposed MPI operations to
support the ST communication strategy. The proposed ST MPI operations follow
semantics similar to the point-to-point (P2P) two-sided messaging. These ST
operations are used to explore the potential performance of the new
communication strategy with minimal network resource utilization. Also, these
proposed operations are designed to be used incrementally within existing MPI
applications.

\subsection{MPI Queues for P2P ST}
\label{src:mpi-queues-p2p-st}

\textit{MPIX\_Queue} is a new data object created to support ST MPI operations
using P2P messaging semantics. The proposed \textit{MPIX\_Queue} object allows
users to map a user-defined GPU stream handle to the MPI runtime, and it
enables batching of operations\footnote{Batching of operations allows users
to trigger execution of multiple enqueued MPI operations using a single
trigger operation. Extracting deferred semantics on enqueued operations and
triggering those deferred operations are provided as part of the proposed
functions in Section~\ref{src:enqueue}.}.
\textit{MPIX\_Create\_queue} and \textit{MPIX\_Free\_queue} are the two new MPI
operations to create and destroy an MPI queue object, respectively. The function
prototypes of these operations are provided in Fig.~\ref{fig:queue-proto}. These
are local operations performed without interaction with any other process.

\begin{figure}[!ht]
\begin{minipage}{\linewidth}
\lstset{
    language=C,
    basicstyle=\tt\small,
    frame = single,
    morekeywords={MPIX_Queue},
}
\begin{lstlisting}
int MPIX_Create_queue(IN void *stream,
                      OUT MPIX_Queue *queue);
int MPIX_Free_queue(IN MPIX_Queue queue);
\end{lstlisting}
\end{minipage}
\caption{Function prototypes for creating and destroying \textit{MPIX\_Queue}
objects.}
\label{fig:queue-proto}
\end{figure}

The \textit{MPIX\_Create\_queue} operation creates a new \textit{MPIX\_Queue}
object, with the GPU stream handle passed by the user as input. Similarly,
the \textit{MPIX\_Free\_queue} allows the implementation to release an already
created \textit{MPIX\_Queue} object. Note that the free operation is used only
to release any internal resource maintained by the queue object, and it is the
responsibility of the users to wait for any associated ST operation to complete
before freeing the queue object.

\subsection{MPI Enqueue Operations for P2P ST}
\label{src:enqueue}

Set of enqueue operations, \textit{MPIX\_Enqueue\_send} and
\textit{MPIX\_Enqueue\_recv}, introduced in this section performs the actual GPU
stream-based data movement operations. Along with these enqueue operations,
\textit{MPIX\_Enqueue\_start} and \textit{MPIX\_Enqueue\_wait} operations are
introduced to trigger and wait for completion of the enqueued data movement
operations, respectively. The function prototypes of these operations are
provided in Fig.~\ref{fig:enqueue-proto} and their semantics are enumerated
below.

\begin{figure}[!ht]
\begin{minipage}{\linewidth}
\lstset{
    language=C,
    basicstyle=\tt\small,
    frame = single,
    morekeywords={MPI_Datatype, MPI_Comm, MPIX_Queue, MPI_Request},
}
\begin{lstlisting}
int MPIX_Enqueue_send(const void *buf,
    int count, MPI_Datatype datatype, int dest,
    int tag, MPI_Comm comm, MPIX_Queue queue,
    MPI_Request *request);

int MPIX_Enqueue_recv(void *buf, int count,
    MPI_Datatype datatype, int source, int tag,
    MPI_Comm comm, MPIX_Queue queue,
    MPI_Request *request);

int MPIX_Enqueue_start(const MPIX_Queue queue);

int MPIX_Enqueue_wait(const MPIX_Queue queue);
\end{lstlisting}
\end{minipage}
\caption{Function prototypes for enqueuing P2P ST communication.}
\label{fig:enqueue-proto}
\end{figure}

\begin{enumerate}
    \item All proposed new ST-based MPI communication operations are enqueued to
    a particular \textit{MPIX\_Queue} object, and these operations are executed
    in FIFO order. Execution of these operations is asynchronous with respect
    to the host process running on the CPU.
    \item \textit{MPIX\_Enqueue\_send} and \textit{MPIX\_Enqueue\_recv} are
    examples of ST MPI communication operations. These operations create
    internal communication descriptors that correspond to MPI data-movement
    operations, and these descriptors are enqueued into the \textit{MPIX\_Queue}
    object.
    \item The \textit{MPIX\_Enqueue\_start} function allows an application
    process to specify when the enqueued stream-based MPI communication
    operations must be executed by the GPU CP. All previously enqueued
    stream-based MPI communication operations on the \textit{MPIX\_Queue} object
    are executed in batch by a single start operation. It is not necessary to
    create a start operation per enqueued stream-based MPI communication
    operation. A stream memory \textit{writeValue} is enqueued internally as
    part of the enqueue start operation.
    \item Similarly, \textit{MPIX\_Enqueue\_wait} allows an application
    process to define when the GPU CP must wait for the completion of all
    previously executed stream-based communication operations, and a stream
    memory \textit{waitValue} operation is enqueued internally as part of this
    operation.
\end{enumerate}

\subsubsection{ST Execution Sequence}
\label{src:st-exec}

\begin{figure}[ht]
  \includegraphics[width=\linewidth]{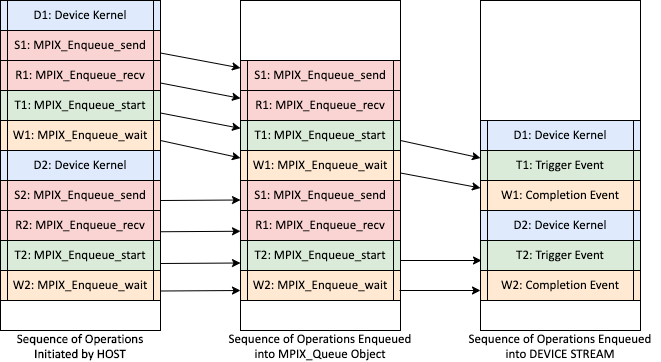}
  \caption{Illustrating a sequence of ST MPI operations and their execution.}
  \label{fig:st-exec-seq}
\end{figure}

Fig.~\ref{fig:st-exec-seq} demonstrates a sequence of ST operations executed in
a simple MPI application. The first queue in Fig.~\ref{fig:st-exec-seq} shows
the sequence of operations executed by the application process executing on the
CPU, while the second and third queues show the list of enqueued operations in
the \textit{MPIX\_Queue} object and the corresponding GPU device stream,
respectively.

The application process directly enqueues the device kernels (D1 and D2) to be
executed by the GPU stream. The \textit{MPIX\_Queue} object is not relevant in
the execution of compute kernels that are offloaded to the GPU.

When an application process calls \textit{MPIX\_Enqueue\_send} (S1, S2) and
\textit{MPIX\_Enqueue\_recv} (R1, R2) operations, they are enqueued in the
\textit{MPIX\_Queue} object. The \textit{MPIX\_Enqueue\_send} and
\textit{MPIX\_Enqueue\_recv} operations return immediately. The enqueue
operations create the necessary communication descriptors for executing the ST
send and receive operations in a deferred manner.

\begin{figure*}[t]
\begin{minipage}{\textwidth}
\lstset{
    language=C,
    basicstyle=\tt\small,
    frame = single,
    morekeywords={MPIX_Queue, hipStream_t, hipStreamCreateWithFlags,
    MPIX_Create_queue, MPIX_Free_queue, hipStreamDestroy, hipStreamSynchronize,
    MPIX_Enqueue_send, MPIX_Enqueue_recv, MPIX_Enqueue_start, MPIX_Enqueue_wait},
}
\begin{lstlisting}
MPIX_Queue queue;
hipStream_t stream;

/* create a GPU stream object and use it to create an MPIX_Queue object */
hipStreamCreateWithFlags(&stream, hipStreamNonBlocking);
MPIX_Create_queue(MPI_COMM_WORLD_DUP, (void *)stream, &queue);

if (my_rank == 0) {
    launch_device_compute_kernel(src_buf1, src_buf2, src_buf3, src_buf4, stream);

    MPIX_Enqueue_send(src_buf1, SIZE, MPI_INT, 1, 123, queue, &sreq[0]);
    MPIX_Enqueue_send(src_buf2, SIZE, MPI_INT, 1, 126, queue, &sreq[1]);
    MPIX_Enqueue_send(src_buf3, SIZE, MPI_INT, 1, 125, queue, &sreq[2]);
    MPIX_Enqueue_send(src_buf4, SIZE, MPI_INT, 1, 124, queue, &sreq[3]);

    MPIX_Enqueue_start(queue); /* Enqueue_start enables triggering of all prior send ops */
    MPIX_Enqueue_wait(queue);  /* wait blocks only the current GPU stream */
} else if (my_rank == 1) {
    MPIX_Enqueue_recv(dst_buf1, SIZE, MPI_INT, 0, 123, queue, &rreq[0]);
    MPIX_Enqueue_recv(dst_buf2, SIZE, MPI_INT, 0, 126, queue, &rreq[1]);
    MPIX_Enqueue_recv(dst_buf3, SIZE, MPI_INT, 0, 125, queue, &rreq[2]);
    MPIX_Enqueue_recv(dst_buf4, SIZE, MPI_INT, 0, 124, queue, &rreq[3]);

    MPIX_Enqueue_start(queue);
    MPIX_Enqueue_wait(queue);

    launch_device_compute_kernel(dst_buf1, dst_buf2, dst_buf3, dst_buf4, stream);
}
hipStreamSynchronize(stream);/* wait for all operations on stream to complete */

MPIX_Free_queue(queue);
hipStreamDestroy(stream);
\end{lstlisting}
\caption{Example for batched ST communication operation.
Illustrate using a single \textit{MPIX\_Enqueue\_start} operation to trigger the
execution of multiple ST send and receive operations.}
\label{fig:example1}
\end{minipage}
\end{figure*}

\textit{MPIX\_Enqueue\_start} internally enqueues a \textit{writeValue}
operation to the specified GPU stream. The execution of the \textit{writeValue}
operation enqueued as part of \textit{T1: MPIX\_Enqueue\_start} triggers the
execution of S1 and R1. Similarly, the execution of the \textit{writeValue}
operation enqueued as part of \textit{T2: MPIX\_Enqueue\_start} triggers the
execution of S2 and R2.

When the ST completion operations (W1 and W2) are called by the
application process using the \textit{MPIX\_Enqueue\_wait} operations, the MPI
implementation enqueues these operations into the \textit{MPIX\_Queue} object,
and the \textit{MPIX\_Enqueue\_wait} operations return immediately. For each
\textit{MPIX\_Enqueue\_wait} operation, the MPI implementation also enqueues
a \textit{waitValue} operation to the corresponding GPU stream. The wait
operation allows the GPU CP to wait for all previously started
stream-based MPI communication operations to complete before working on other
 enqueued stream-based MPI operations and device compute kernels.

\subsubsection{Non-blocking ST Semantics}
\label{src:nbi-st}

The proposed enqueue operations are non-blocking with respect to the
application host process. The non-blocking semantics of the proposed enqueue
operations eliminates the expensive synchronization points between an
application process and its GPU device. The summary of the non-blocking
semantics of the new proposed operations is as follows.

\begin{enumerate}
    \item All ST enqueue operations return after enqueuing them into
    the \textit{MPIX\_Queue} object. The execution of these operations is
    deferred until the GPU CP triggers the execution using the enqueued
    \textit{writeValue} operation as part of the \textit{MPIX\_Enqueue\_start}.
    \item Once enqueued, the GPU device kernels can make changes to these
    buffers until the execution of the GPU stream memory write operations in
    stream order.
    \item An application process can synchronize with the GPU device to identify
    the completion of previously enqueued stream operations. This is in addition
    to the \textit{waitValue} operation enqueued as part of the
    \textit{MPIX\_Enqueue\_wait} operation. When a process enqueues a
    \textit{waitValue} operation, it is not synchronizing with the GPU. Instead
    a wait command is enqueued that gets executed by the GPU CP.
    \item An application process can also call \textit{MPI\_Wait} or
    \textit{MPI\_Waitall} to ensure the completion of ST communication
    operations and perform any related cleanup activities. These host-based
    synchronization operations have blocking semantics for an application
    process.
\end{enumerate}

\subsection{ST Usage Model}
\label{src:st-examples}
This section provides a simple example in Fig.~\ref{fig:example1} to show the
usage of the proposed ST operations. In this example, each process creates
an \textit{MPIX\_Queue} object on a given GPU device stream. The MPI process
with \textit{my\_rank} value 0 in \textit{MPI\_COMM\_WORLD\_DUP} first launches
a compute kernel to the GPU. Since the kernel launch operation is non-blocking,
this process can return immediately to enqueue further stream-based send
operations to the \textit{MPIX\_Queue} and the corresponding GPU stream objects.
This process next calls \textit{MPIX\_Enqueue\_start}, which appends the
\textit{writeValue} to the corresponding GPU stream. Finally, this
process calls \textit{MPIX\_Enqueue\_wait}, which appends a
\textit{waitValue} to the GPU stream. Similarly, the MPI process with
\textit{my\_rank} value 1 on \textit{MPI\_COMM\_WORLD\_DUP} performs matching
receive operations on the same communicator by calling
\textit{MPIX\_Enqueue\_recv}, \textit{MPIX\_Enqueue\_start}, and
\textit{MPIX\_Enqueue\_wait} operations on the specified GPU stream.

In this example, for the MPI process with \textit{my\_rank} value 0, the enqueued
stream-based sends are guaranteed to be executed only after the completion of
the device kernel execution. The host process enqueues the kernel and the
stream-based sends and start operations on a given stream handle and returns
immediately. It is the GPU CP that executes these operations in
FIFO order.

Similarly, for the MPI process with \textit{my\_rank} value 1, the enqueued wait
operation guarantees that the device kernel is not executed until the enqueued
stream-based receive operations have completed execution. This also ensures that
the \textit{dst\_buf} memory regions contain the result of the stream-based
receive operations for the subsequent GPU kernel enqueued for execution.

\subsection{ST Extended Usage}
\label{src:st-extended-usage}
This section enumerates the extended usage semantics of the proposed ST
operations:
\begin{enumerate}
    \item The proposed MPI enqueue operations are fully compatible with existing
    MPI P2P communication operations. For example, as long as the message
    matching requirements are satisfied, the proposed API allows the use of
    existing \textit{MPI\_Irecv} along with the \textit{MPIX\_Enqueue\_send}.
    \item The proposed ST operations do not support wildcarding. Usage of
    \textit{MPI\_ANY\_SOURCE} and \textit{MPI\_ANY\_TAG} is restricted.
\end{enumerate}

\section{Implementation Details}
\label{src:implementation-details}

This section provides a brief overview of the MPI implementation details in
supporting the proposed ST-based MPI communication operations. Implementation
details are divided into two different sections, providing an overview of
the inter-node and intra-node implementations.

\subsection{Inter-node ST MPI Implementation}
\label{src:inter-node-implementation}
Inter-node ST send operations are executed using the triggered operations
supported by Slingshot 11. This feature is exposed using the DWQ features
supported in Libfabric interface.

When the \textit{MPIX\_Queue} object is created using the
\textit{MPIX\_Create\_queue} operation, each MPI process opens two Libfabric
counters that are associated with hardware counters in the Slingshot 11 NIC.
This implementation uses one hardware counter as the trigger counter and the
other as the completion counter for all ST-based send and receive
operations enqueued on this \textit{MPIX\_Queue} object.

\subsubsection{Inter-node ST MPI Send Implementation}
\label{src:inter-node-st-send}
During \textit{MPIX\_Enqueue\_send} operation, a DWQ-based send operation with
the hardware counters associated with the corresponding \textit{MPIX\_Queue}
object is created. Appropriate trigger threshold values are set per enqueued
send operation. As per the supported semantics of the DWQ operations, these
enqueued send operations are not executed immediately. They wait for the
trigger condition on the associated hardware trigger counter to be satisfied.

With \textit{MPIX\_Enqueue\_start} operation, a trigger event using the
\textit{writeValue} operation is created to satisfy the trigger condition. The
write operation to the associated hardware counter from the \textit{writeValue}
operation acts as a trigger to any previously enqueued triggered operations.

\subsubsection{Inter-node ST MPI Receive Implementation}
\label{src:inter-node-st-recv}
The implementation of the \textit{MPIX\_Enqueue\_recv} operation is similar to
the \textit{MPIX\_Enqueue\_send} operation. But the main difference is that, as
the Slingshot 11 NIC does not provide support for triggered receives, the
deferred execution semantics of the triggered receives are emulated using a
asynchronous progress thread running on the CPU. The progress thread monitors
the associated hardware trigger and completion counter, and interprets their
state to decide when to execute the enqueued receive operation and wait for
its completion.

While the inter-node send operations are fully offloaded to the network, the
deferred execution semantics of the inter-node receive operations are emulated
using a progress thread.

\subsection{Intra-node ST MPI Implementation}
\label{src:intra-node-implementation}
For both intra-node ST-based send and receive operations, there are no known
peer-to-peer options available to provide the required deferred work semantics
(specifically the MPI messaging matching~\cite{bmm,hw_mm,osti_hw_mm,coll_mm}
required as part of the message passing semantics). As mentioned in
Section~\ref{src:inter-node-st-recv}, where the inter-node ST-based receive
operations are emulated using a progress thread, all intra-node ST-based send
and receive operations are also emulated using the same progress thread.

It is to note that the described implementation is possible, as the proposed
ST operations do not support wildcarding with \textit{MPI\_ANY\_SOURCE} and
\textit{MPI\_ANY\_TAG} options. With this semantic, both inter-node and
intra-node ST traffic are easily separable. And, the progress thread-based
emulation would allow the asynchronous progress thread to get involved in the
message matching and data movement operations by monitoring the memory buffer
updated by the stream memory operations mentioned in
Section~\ref{src:stream-memory-operations}. While the progress thread handles
the control path for the intra-node operations, the GPU vendor supported DMA
engines are used for the data paths of these operations.

\section{Performance Analysis}
\label{src:perf}
In this section, the Faces microbenchmark is used to evaluate the performance
of the proposed ST-based MPI P2P communication operations.

\subsection{Application Overview}
\label{src:app-overview}

The Faces microbenchmark is based on the nearest-neighbor communication pattern
from the CORAL-2~\cite{coral2-bm} Nekbone~\cite{nekbone-bm} benchmark. The timed
loop in Faces performs the following operations in the baseline HIP
implementation.

\begin{enumerate}
    \item Pre-post non-blocking MPI receives from up to 26 neighbors;
    \item Launch kernels to copy into contiguous MPI buffers from faces, edges,
    and corners of spectral elements on the surface of the 3D local block, all
    in GPU memory;
    \item Initiate non-blocking MPI sends to all neighbors;
    \item Launch a kernel to perform the sum operation of all faces of spectral
    elements that are inside the local block;
    \item Wait to receive messages from neighbors; and
    \item Launch kernels to add the received messages to the faces, edges, and
    corners of spectral elements on the surface of the local block.
\end{enumerate}

Faces has three nested loops that perform the following operations.
\begin{enumerate}
    \item \textit{Outer loop}: Allocate MPI buffers, run loops, and de-allocate
    MPI buffers;
    \item \textit{Middle loop}: Initialize the values of the spectral elements
    and run the inner loop; and
    \item \textit{Inner loop}: Run the communication steps listed above and
    accumulate the wall-clock runtime.
\end{enumerate}
Faces confirms correct results by comparing against a reference CPU-only
implementation.

\subsection{Test Details}
\label{src:test-details}
For this ST performance analysis, the baseline HIP implementation of the Faces
microbenchmark is compared against the ST modified variant of the same. The
baseline HIP implementation uses GPU-attached memory regions and attempts to
overlap MPI communication with computation for the inner domain.

The ST implementation of Faces replaces \textit{MPI\_Isend} calls with
\textit{MPIX\_Enqueue\_send} calls followed by \textit{MPIX\_Enqueue\_start}.
These changes eliminate the need for host-device synchronization before the
sends. The implementation also replaces the synchronous call to
\textit{MPI\_Waitall} for the sends with the host-asynchronous call
\textit{MPIX\_Enqueue\_wait}.

For receive operations, Faces can use standard \textit{MPI\_Irecv}
operations with appropriate buffer management techniques, or it can use the
proposed ST receive operations, \textit{MPIX\_Enqueue\_recv}. For this study,
the modified version of Faces uses standard \textit{MPI\_Irecv} operations with
double buffering techniques. This is an intentional implementation choice
because the current Slingshot 11 NIC does not support deferred receive
operations. Pre-posting the receive operations eliminates the need for
MPI’s progress threads to respond to the trigger event (from
\textit{hipStreamWriteValue64}) and post the receive descriptor to the NIC.
Thus, the receive side of the Faces implementation uses standard
\textit{MPI\_Irecv} calls.

For all tests, the following loop configurations were used:
10 outer loop $\times$ 100 middle loop $\times$ 100 inner loop. 5 different runs
of the different variants of the tests were performed, and the average of the
results are reported in this analysis.

\subsection{Initial Performance Evaluation}
\label{src:initial-perf}
The results of 1D Faces runs using the baseline and ST variants are reported in
this section. A $64\times 1\times 1$ (1D) MPI distribution is used for the
analysis. This test uses 64 application processes (MPI ranks) distributed across
8 heterogeneous nodes, with 8 ranks per node. On each node, a one-to-one
mapping between MPI rank and GPU devices is enforced. On each node,
8 GPU devices (4 sockets) are attached to a single CPU. The heterogeneous node
architecture is similar to the node architecture available in the Frontier
exascale supercomputer~\cite{frontier}.

\begin{figure}[ht]
  \includegraphics[width=\linewidth]{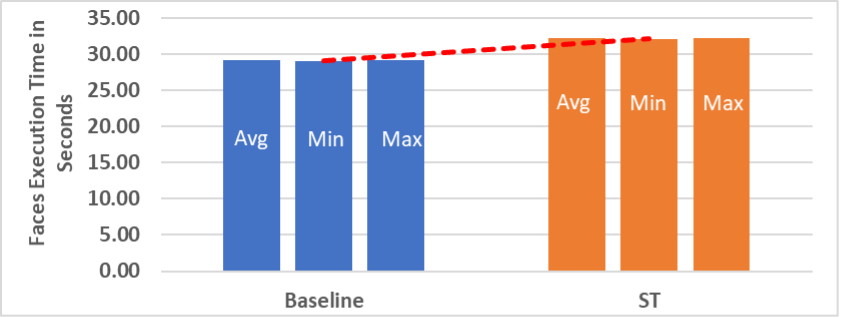}
  \caption{Faces execution time in seconds for the baseline and ST
  variants of the benchmark (8 nodes with 8 MPI processes per node,
  $64\times 1\times 1$ 1D configuration).}
  \label{fig:initial-perf}
\end{figure}

Fig.~\ref{fig:initial-perf} represents the results of the analysis.
It shows the average, minimum, and maximum overall
execution time of the Faces microbenchmark, comparing the baseline version
against the modified ST variant.

On average, the ST variant of the Faces execution is around 10\% slower than the
baseline version. In the remainder of the performance analysis sections,
the performance impact of the various components in the ST design is
investigated. Multiple intra-node and inter-node tests are used to understand
the various factors contributing to the performance overheads measured in this
multi-node analysis.

\subsection{Impact of Progress Threads}
\label{src:progress-threads}
To understand the impact of progress thread usage in the ST implementation, the
Faces microbenchmark is tested on a single node. For this analysis, the 1D
distribution of application processes is $8\times 1\times 1$.

\begin{figure}[ht]
  \includegraphics[width=\linewidth]{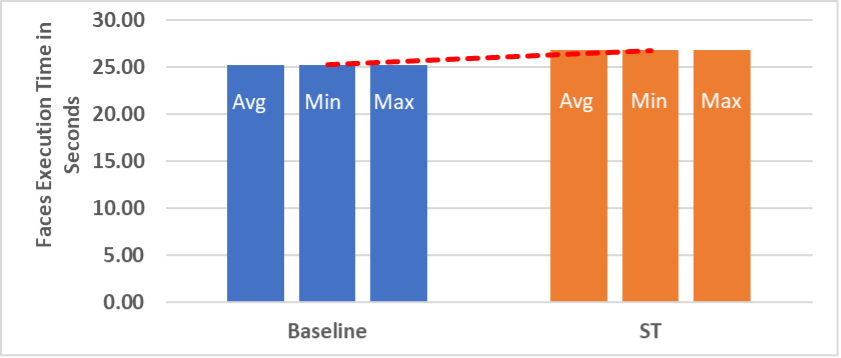}
  \caption{Faces execution time in seconds for the baseline and ST
  variants of the benchmark (1 node with 8 MPI processes per node,
  $8\times 1\times 1$ 1D configuration).}
  \label{fig:progress-threads}
\end{figure}

All eight MPI ranks are collocated on a single node, with each MPI rank using a
different GPU on the node. Fig.~\ref{fig:progress-threads} represents the
results of the analysis.

With the $8\times 1\times 1$ 1D configuration and all ranks placed on a single
node, MPI messaging operations for the baseline version involve the use of
\textit{ROCr}~\cite{rocr} (user-mode API interfaces and libraries necessary
for the host applications to launch compute kernels on AMD GPU devices)
IPC (Inter-Process Communication) for large payloads and a non-temporal memcpy
implementation for small payloads. While the data transfers for the ST variant
also incur similar intra-node transfers, a progress thread is employed to
perform these operations to satisfy the ST deferred execution semantics. The
details of the progress thread usage in the ST implementation design are
described in Section~\ref{src:intra-node-implementation}.

On average, the ST variant of the Faces execution is around 4\% slower than the
baseline version. This is observed even with a dedicated hardware
thread on the AMD CPU for each MPI progress thread. This demonstrates the
negative impact of using a progress thread per MPI process to emulate the
triggered operation semantics to implement the ST semantics.

\subsection{Impact of Network Offload}
\label{src:network-offload}
\begin{figure}[ht]
  \includegraphics[width=\linewidth]{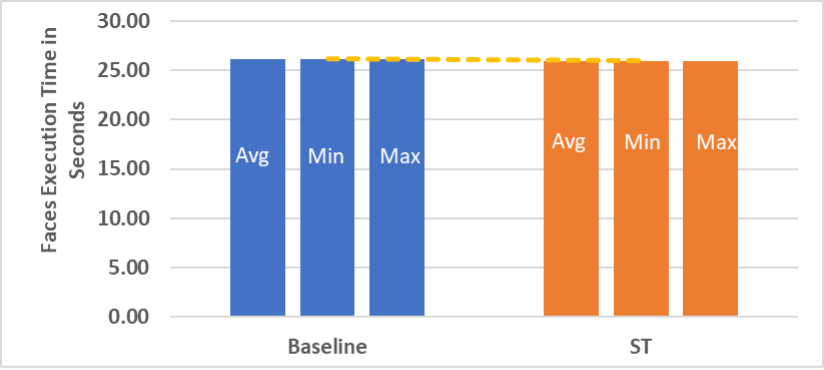}
  \caption{Faces execution time in seconds for the baseline and ST
  variants of the benchmark (8 nodes with 1 MPI processes per node,
  $8\times 1\times 1$ 1D configuration).}
  \label{fig:network-offload-1d}
\end{figure}

To understand the impact of using HPE Slingshot 11 hardware-based deferred
execution operations in the ST implementation, the Faces microbenchmark is
tested across 8 nodes with 1 MPI rank per node. For this analysis, the 1D
distribution of the ranks is $8\times 1\times 1$. On each node, a given MPI
rank uses a single GPU device, and the Slingshot 11 NIC is co-located on the same
GPU module. Fig.~\ref{fig:network-offload-1d} represents the results of this
analysis.

With the $8\times 1\times 1$ 1D configuration and a single MPI rank per node,
all MPI messaging operations for the baseline version involve inter-node data
transfers through the HPE Slingshot-11 network. While the data transfers for the
ST variant also incur similar inter-node data transfers, HPE Slingshot 11
hardware-based deferred send operations and hardware counters are employed to
perform these operations to satisfy the ST semantics. The details of the
implementation are provided in Section~\ref{src:inter-node-st-send}.

\begin{figure}[ht]
  \includegraphics[width=\linewidth]{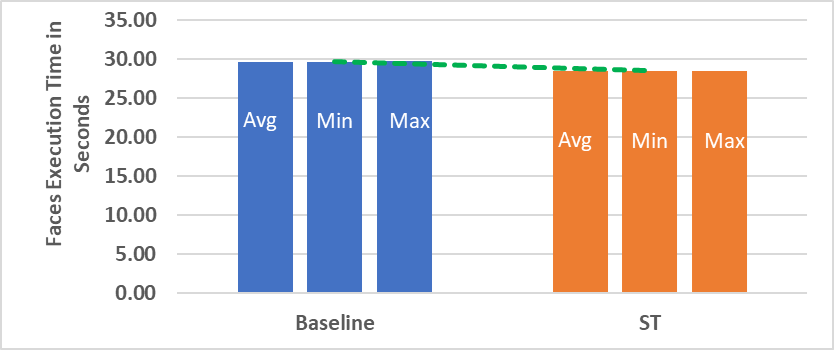}
  \caption{Faces execution time in seconds for the baseline and ST
  variants of the benchmarks (8 nodes with 1 MPI processes per node,
  $2\times 2\times 2$ 3D configuration).}
  \label{fig:network-offload-3d}
\end{figure}

In this test, on average the ST variant shows similar performance to
the baseline Faces variant. This shows that offloading significant components of
the ST operations to the HPE Slingshot 11 NIC offers better performance when
compared to the use of a progress thread per MPI process to implement the
intra-node ST operations. To confirm this, the Faces benchmark is run with a
$2\times 2\times 2$ 3D distribution. The 3D distribution increases the number of
MPI messages generated per rank and the number of target ranks per process. The
number of MPI ranks and the number of nodes used for this test stays the same;
only the Faces distribution is changed for this analysis. The results of this
analysis are shown in Fig.~\ref{fig:network-offload-3d}.

With the change in Faces distribution, the performance clearly shows the benefit
of using NIC offloading for the ST implementation. On average, the ST variant of
the Faces execution as shown in Fig.~\ref{fig:network-offload-1d} is around 4\%
better than the baseline version. This experiment demonstrates the benefits of
using hardware-based deferred execution operations in the HPE Slingshot 11 NIC
to implement ST operations.

Even though the HPE Slingshot 11 NIC offers hardware support for deferred
execution operations and completion counters, there are scenarios where the MPI
implementation relies on the progress thread for inter-node ST operations.
However, the number of CPU cycles required to mitigate some of the missing
hardware capabilities is much smaller when compared to those needed for the
intra-node ST implementation. Specifically, CPU cycles are needed
for emulating deferred receives and for handling completion counter updates in
specific inter-node scenarios, such as the rendezvous protocol. The NIC handles
the entire progression of the rendezvous protocol and the data movement operations.
In summary, hardware capabilities in the HPE Slingshot 11 NIC can offer performance
benefits for inter-node ST operations, even though the current implementation
requires CPU cycles for some scenarios.

\subsection{Impact of Tuned Stream Memory Operations}
\label{src:stream-memory}
In this test, the impact of using HIP stream memory operations as described in
Section~\ref{src:stream-memory-operations} is analyzed. For this test, the
$2\times 2\times 2$ 3D distribution of Faces execution as described in
Section~\ref{src:network-offload} is extended.

Here 8 MPI ranks are distributed across 8 nodes, with 1 MPI rank per
node and each rank uses a single GPU. The $2\times 2\times 2$ 3D distribution is
employed to show the real benefits of the ST implementation.

As described in Section~\ref{src:network-offload}, the ST implementation that
uses the HIP stream memory operation (\textit{hipStreamWriteValue64} and
\textit{hipStreamWaitValue64}) shows 4\% better performance when compared to the
baseline version. The performance improvements are attributed to the use of
HPE Slingshot 11 hardware-based deferred execution operations in the ST
inter-node data transfer implementation.

\begin{figure}[ht]
  \includegraphics[width=\linewidth]{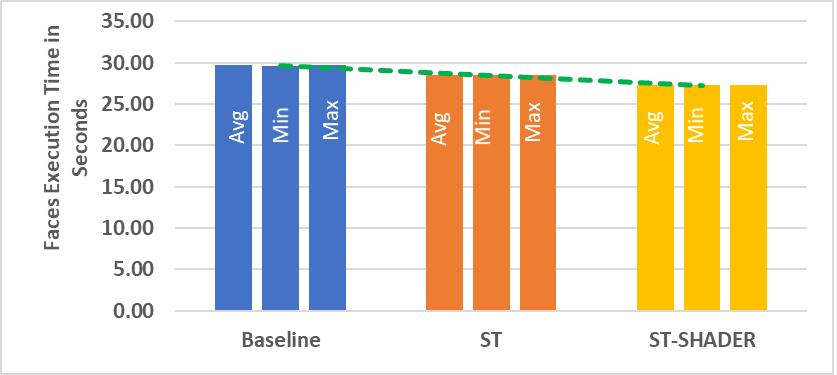}
  \caption{Faces execution time in seconds for the baseline, ST
  variant with HIP stream memory operations, and ST implementation
  with hand-coded shader kernels (8 nodes with 1 MPI rank per node using
  $2\times 2\times 2$ 3D configuration).}
  \label{fig:stream-memory}
\end{figure}

For this test, the HIP stream memory operations are swapped with hand-coded
shaders that perform a set of operations to satisfy the semantics of the
\textit{hipStreamWriteValue64} and \textit{hipStreamWaitValue64} operations. The
ST-shader variants of Faces show the best performance when
compared to the baseline, as well as the regular ST variant with HIP stream
memory operations. The results of this analysis are shown in
Fig.~\ref{fig:stream-memory}. The ST-shader version shows 8\% better performance
than the baseline version.

This analysis shows the need for further tuning the HIP stream memory operations
to suite the need for ST MPI operations.

\subsection{Performance Inference}
\label{src:inference}

In brief, performance analysis comparing the baseline and ST variants of the Faces
microbenchmark determines the following.

\begin{enumerate}
    \item On average, experiments with the ST variant of Faces with multiple
    nodes and multiple processes per node show a significant drop in performance
    when compared to the baseline variant of Faces. Specifically, an 8-node
    experiment with 8 MPI processes per node experiment shows a 10\% drop in
    overall performance when using the ST implementation. This drop in
    performance is attributed to the use of progress threads to emulate the
    deferred execution semantics needed for intra-node data transfers.
    \item Running the ST variant of the Faces benchmark across multiple nodes
    with a single application process per node shows performance improvements
    when compared to the baseline variant of Faces. Specifically, an 8-node
    experiment with the ST version of Faces with one MPI rank per node shows 4\%
    improvement in overall Faces execution time. This performance improvement
    can be attributed to the use of hardware capabilities in the
    HPE Slingshot 11 NIC in the ST implementation.
    \item While the performance benefit from exploiting the hardware-based deferred
    execution is an interesting data point, not all applications can limit the
    amount of intra-node communication to mimic the layout that provides
    performance benefits. For the baseline implementation, nearest-neighbor
    codes benefit from rank ordering that puts communicating neighbors on the
    same compute nodes. But for ST, a rank order that keeps neighbors on
    separate nodes shows a greater improvement over the standard implementation,
    but the absolute performance relative to node-localized orders is likely
    lower. Further study of this space may be appropriate.
    \item The 4\% performance improvement while using HPE Slingshot 11
    deferred execution operations can be improved to 8\% by employing
    hand-coded shader kernels replacing the HIP stream memory
    operations, \textit{hipStreamWriteValue64} and \textit{hipStreamWaitValue64}.
    This shows the need for further performance tuning of targeted stream
    memory operations.
\end{enumerate}

\section{Related Work}
\label{src:relate}
To the best of our knowledge, little research has been performed to explore
options for introducing GPU stream-awareness into message-passing
programming models with deferred execution semantics~\cite{tops-portal}. The
triggered operations functionality supported by the HPE Slingshot NIC
(Slingshot 11) is unique in allowing users to enqueue communication events into
the NIC command queue but defer their execution until a trigger event is
observed. Based on our understanding, this is the first research to explore the
usage of a deferred execution programming model to introduce GPU stream-awareness
in MPI.

There is other related research exploring different hardware options to introduce
GPU stream-awareness in different programming models.
\citeauthor{offload-communication-control} explores the basic building blocks
required for offloading communication control logic in GPU accelerated
applications into the GPU device. Similarly to our research, they have explored
hardware offload capabilities using NVIDIA GPUDirect Async~\cite{nvda-gpudirect}
and InfiniBand Connect-IB network adapters\cite{verbs-rdma}.
The \textit{libgdsync}\cite{nvda-gdsync} library was created as part of this
research.

\citeauthor{gpu-verbs} and \citeauthor{gpurdma} explore the possibilities of
GPUs handling the communication and control paths by hosting the verbs
layer~\cite{verbs-rdma} in the GPU. Similarly, \citeauthor{nvshmem} and
\citeauthor{rocshmem} explore the options to host the GPU centric communication
on the GPU using the OpenSHMEM programming model\cite{osm14}. This
research~\cite{gpu-verbs,gpurdma,nvshmem,rocshmem} is mostly done using the
NVIDIA Infiniband Host Channel Adapter.

We conclude that the proposed GPU stream-aware MPI operations with deferred
execution semantics have a potential for performance benefits. Specifically,
the use of hardware offloaded deferred execution operations can benefit in
implementing the proposed MPI operations. The missing peer-to-peer
intra-node data transfer support needs similar offload mechanism for exploiting
full performance benefit from the proposed MPI operations.

\section{Conclusion}
\label{src:conclusion}

In this work, a new communication strategy called \textit{stream-triggered
communication} is proposed to introduce GPU stream-awareness in MPI 2-sided
point-to-point communication operations. The proposed strategy allows
an application to offload both the \textit{control} and \textit{data} paths to
the underlying implementation and hardware components, and it avoids an active
CPU-GPU synchronization at the GPU compute kernel boundaries.

The implementation of the proposed ST-based MPI operations shows that offloading
the inter-node ST-based MPI operations using the triggered operations feature
supported in NICs like HPE Slingshot 11 has shown performance benefits.
Conversely, the usage of progress threads for the intra-node communication to
emulate the required deferred execution semantics is detrimental to performance.
With available systems, it is not yet possible for us to efficiently implement
intra-node two-sided MPI operations with the ST semantics without a progress
thread. The intra-node implementation requires a progress thread per MPI process
to perform the MPI message matching for intra-node communication operations.

Further analysis is required to identify options to fully offload the ST
communication semantics to the NIC hardware to get the maximum performance
benefits from new interfaces.

\section{Acknowledgment and Disclaimer}\label{src:ack}
We would like to thank the HPE MPI (Steve Oyanagi and Arm Patinyasakdikul)
and HPE Libfabric (Ian Ziemba and Charles Fossen) developers involved in
providing the basic features required to implement and evaluate the proposed MPI
stream-aware communication strategy. Also, we would like to thank Duncan Roweth
and Keith Underwood for reviewing our work and providing suggestions for
further improvements. Any opinions, findings, and conclusions or recommendations
expressed in this material are those of the authors and do not necessarily
reflect the views of associated organizations.

\bibliography{master}

\begin{thebibliography}{31}
\providecommand{\natexlab}[1]{#1}
\providecommand{\url}[1]{\texttt{#1}}
\expandafter\ifx\csname urlstyle\endcsname\relax
  \providecommand{\doi}[1]{doi: #1}\else
  \providecommand{\doi}{doi: \begingroup \urlstyle{rm}\Url}\fi

\bibitem[cor()]{coral2-bm}
{CORAL-2 Benchmarks Summary}.
\newblock \url{https://asc.llnl.gov/coral-2-benchmarks}.

\bibitem[cud()]{cuda-stream-memory-ops}
{CUDA Stream Memory Operations}.
\newblock
  \url{https://docs.nvidia.com/cuda/cuda-driver-api/group__CUDA__MEMOP.html}.

\bibitem[fi_()]{fi_trigger}
{Libfabric Deferred Work Queue}.
\newblock \url{https://ofiwg.github.io/libfabric/v1.9.1/man/fi_trigger.3.html}.

\bibitem[fro()]{frontier}
{Frontier, ORNL's Exascale Supercomputer}.
\newblock \url{https://www.olcf.ornl.gov/frontier/}.

\bibitem[hip()]{hip-stream-memory-ops}
{HIP Stream Memory Operations}.
\newblock
  \url{https://github.com/ROCm-Developer-Tools/HIP/blob/develop/docs/markdown/hip_programming_guide.md#hip-stream-memory-operations}.

\bibitem[nek()]{nekbone-bm}
{CORAL-2 Benchmarks Summary - Nekbone}.
\newblock
  \url{https://asc.llnl.gov/sites/asc/files/2020-06/Nekbone_Summary_v2.3.4.1.pdf}.

\bibitem[nvd({\natexlab{a}})]{nvda-gdsync}
{NVIDIA GPUDirect libgdsync}.
\newblock \url{https://github.com/gpudirect/libgdsync}, {\natexlab{a}}.

\bibitem[nvd({\natexlab{b}})]{nvda-gpudirect}
{NVIDIA GPUDirect family}.
\newblock \url{https://developer.nvidia.com/gpudirect}, {\natexlab{b}}.

\bibitem[roc()]{rocr}
{AMD ROCr-Runtime Manpage and Guide}.
\newblock
  \url{https://rocmdocs.amd.com/en/latest/Installation_Guide/ROCR-Runtime.html}.

\bibitem[sli()]{slingshot2}
{HPE Slingshot Interconnect}.
\newblock
  \url{https://www.hpe.com/in/en/compute/hpc/slingshot-interconnect.html}.

\bibitem[ver()]{verbs-rdma}
{IB Verbs RDMA programming guide}.
\newblock
  \url{https://docs.nvidia.com/networking/display/RDMAAwareProgrammingv17/RDMA+Aware+Networks+Programming+User+Manual}.

\bibitem[cud(2013)]{cuda-stream}
{NVIDIA CUDA C/C++ Streams and Concurrency}.
\newblock
  \url{http://on-demand.gputechconf.com/gtc-express/2011/presentations/StreamsAndConcurrencyWebinar.pdf},
  2013.

\bibitem[osm(2017)]{osm14}
{OpenSHMEM standard version-1.4}.
\newblock
  \url{http://openshmem.org/site/sites/default/site_files/OpenSHMEM-1.4.pdf},
  2017.

\bibitem[sli(2022)]{slingshot1}
{Cray's Slingshot Interconnect is at the Heart of HPE's HPC and AI Ambitions}.
\newblock \url{https://tinyurl.com/22rt7utz}, 2022.

\bibitem[Agostini et~al.(2017)Agostini, Rossetti, and
  Potluri]{offload-communication-control}
E.~Agostini, D.~Rossetti, and S.~Potluri.
\newblock {Offloading Communication Control Logic in GPU Accelerated
  Applications}.
\newblock In \emph{2017 17th IEEE/ACM International Symposium on Cluster, Cloud
  and Grid Computing (CCGRID)}, pages 248--257, 2017.
\newblock \doi{10.1109/CCGRID.2017.29}.

\bibitem[Barrett et~al.(2011)Barrett, Brightwell, Hemmert, Wheeler, and
  Underwood]{tops-portal}
B.~W. Barrett, R.~Brightwell, K.~S. Hemmert, K.~B. Wheeler, and K.~D.
  Underwood.
\newblock {Using Triggered Operations to Offload Rendezvous Messages}.
\newblock In \emph{Proceedings of the 18th European MPI Users' Group Conference
  on Recent Advances in the Message Passing Interface}, EuroMPI'11, 2011.

\bibitem[Daoud et~al.(2016)Daoud, Watad, and Silberstein]{gpurdma}
F.~Daoud, A.~Watad, and M.~Silberstein.
\newblock {GPUrdma: GPU-Side Library for High Performance Networking from GPU
  Kernels}.
\newblock In \emph{Proceedings of the 6th International Workshop on Runtime and
  Operating Systems for Supercomputers}, ROSS '16. Association for Computing
  Machinery, 2016.
\newblock ISBN 9781450343879.

\bibitem[De~Sensi et~al.(2020)De~Sensi, Di~Girolamo, McMahon, Roweth, and
  Hoefler]{slingshot}
D.~De~Sensi, S.~Di~Girolamo, K.~H. McMahon, D.~Roweth, and T.~Hoefler.
\newblock {An In-Depth Analysis of the Slingshot Interconnect}.
\newblock In \emph{SC20: International Conference for High Performance
  Computing, Networking, Storage and Analysis}, 2020.

\bibitem[Ferreira et~al.(2019)Ferreira, Grant, Levenhagen, Levy, and
  Groves]{osti_hw_mm}
K.~Ferreira, R.~E. Grant, M.~J. Levenhagen, S.~Levy, and T.~Groves.
\newblock {Hardware MPI message matching: Insights into MPI matching behavior
  to inform design: Hardware MPI message matching}.
\newblock \emph{Concurrency and Computation. Practice and Experience}, 32,
  2019.

\bibitem[Forum(1994)]{mpi}
M.~P. Forum.
\newblock {MPI: A Message-Passing Interface Standard}.
\newblock Technical report, 1994.

\bibitem[Ghazimirsaeed et~al.(2018)Ghazimirsaeed, Grant, and Afsahi]{coll_mm}
S.~M. Ghazimirsaeed, R.~E. Grant, and A.~Afsahi.
\newblock {A Dedicated Message Matching Mechanism for Collective
  Communications}.
\newblock In \emph{Proceedings of the 47th International Conference on Parallel
  Processing Companion}, ICPP '18. Association for Computing Machinery, 2018.

\bibitem[Groves et~al.(2021)Groves, Ravichandrasekaran, Cook, Keen, Trebotich,
  Wright, Alverson, Roweth, and Underwood]{bmm}
T.~Groves, N.~Ravichandrasekaran, B.~Cook, N.~Keen, D.~Trebotich, N.~J. Wright,
  B.~Alverson, D.~Roweth, and K.~Underwood.
\newblock {Not all applications have boring communication patterns: Profiling
  message matching with BMM}.
\newblock \emph{Concurrency and Computation: Practice and Experience}, jun
  2021.
\newblock URL \url{https://doi.org/10.1002%2Fcpe.6380}.

\bibitem[Grun et~al.(2015)Grun, Hefty, Sur, Goodell, Russell, Pritchard, and
  Squyres]{libfabrics}
P.~Grun, S.~Hefty, S.~Sur, D.~Goodell, R.~D. Russell, H.~Pritchard, and J.~M.
  Squyres.
\newblock {A Brief Introduction to the OpenFabrics Interfaces - A New Network
  API for Maximizing High Performance Application Efficiency}.
\newblock Aug 2015.

\bibitem[Hamidouche and LeBeane(2020)]{rocshmem}
K.~Hamidouche and M.~LeBeane.
\newblock \emph{GPU INitiated OPenSHMEM: Correct and Efficient Intra-Kernel
  Networking for DGPUs}.
\newblock Association for Computing Machinery, 2020.

\bibitem[Hemmert et~al.(2007)Hemmert, Underwood, and Rodrigues]{hw_mm}
K.~S. Hemmert, K.~D. Underwood, and A.~Rodrigues.
\newblock {An architecture to perform NIC based MPI matching}.
\newblock In \emph{2007 IEEE International Conference on Cluster Computing},
  2007.
\newblock \doi{10.1109/CLUSTR.2007.4629234}.

\bibitem[Hsu et~al.(2020)Hsu, Imam, Langer, Potluri, and Newburn]{nvshmem}
C.-H. Hsu, N.~Imam, A.~Langer, S.~Potluri, and C.~J. Newburn.
\newblock {An Initial Assessment of NVSHMEM for High Performance Computing}.
\newblock In \emph{2020 IEEE International Parallel and Distributed Processing
  Symposium Workshops (IPDPSW)}, 2020.
\newblock \doi{10.1109/IPDPSW50202.2020.00104}.

\bibitem[Manian et~al.(2019)Manian, Ammar, Ruhela, Chu, Subramoni, and
  Panda]{GPU-aware-intra-node}
K.~V. Manian, A.~A. Ammar, A.~Ruhela, C.-H. Chu, H.~Subramoni, and D.~K. Panda.
\newblock {Characterizing CUDA Unified Memory (UM)-Aware MPI Designs on Modern
  GPU Architectures}.
\newblock In \emph{Proceedings of the 12th Workshop on General Purpose
  Processing Using GPUs}, GPGPU '19, 2019.

\bibitem[Oden et~al.(2014)Oden, Fröning, and Pfreundt]{gpu-verbs}
L.~Oden, H.~Fröning, and F.-J. Pfreundt.
\newblock {Infiniband-Verbs on GPU: A Case Study of Controlling an Infiniband
  Network Device from the GPU}.
\newblock In \emph{2014 IEEE International Parallel Distributed Processing
  Symposium Workshops}, 2014.
\newblock \doi{10.1109/IPDPSW.2014.111}.

\bibitem[Potluri et~al.(2013)Potluri, Hamidouche, Venkatesh, Bureddy, and
  Panda]{GPU-aware-MVAPICH-inter-node}
S.~Potluri, K.~Hamidouche, A.~Venkatesh, D.~Bureddy, and D.~K. Panda.
\newblock {Efficient Inter-node MPI Communication Using GPUDirect RDMA for
  InfiniBand Clusters with NVIDIA GPUs}.
\newblock In \emph{2013 42nd International Conference on Parallel Processing},
  2013.

\bibitem[Wang et~al.(2011)Wang, Potluri, Luo, Singh, Sur, and
  Panda]{GPU-aware-MVAPICH}
H.~Wang, S.~Potluri, M.~Luo, A.~K. Singh, S.~Sur, and D.~K. Panda.
\newblock {MVAPICH2-GPU: optimized GPU to GPU communication for InfiniBand
  clusters}.
\newblock \emph{Computer Science - Research and Development}, 26, 2011.

\bibitem[Wang et~al.(2014)Wang, Potluri, Bureddy, Rosales, and
  Panda]{GPU-aware-MVAPICH-inter-node-2}
H.~Wang, S.~Potluri, D.~Bureddy, C.~Rosales, and D.~K. Panda.
\newblock {GPU-Aware MPI on RDMA-Enabled Clusters: Design, Implementation and
  Evaluation}.
\newblock \emph{IEEE Transactions on Parallel and Distributed Systems}, 25,
  2014.

\end{thebibliography}

\end{document}